# The BNO-LNGS joint measurement of the solar neutrino capture rate in $^{71}$Ga

J. N. Abdurashitov [a], T. J. Bowles [c], C. Cattadori [b,e], B. T. Cleveland [f,*], , S. R. Elliott [c], N. Ferrari [b], V. N. Gavrin [a], S. V. Girin [a], V. V. Gorbachev [a], P. P. Gurkina [a], W. Hampel [d], T. V. Ibragimova [a], F. Kaether [d], A. V. Kalikhov [a], N. G. Khairnasov [a], T. V. Knodel [a], I. N. Mirmov [a], L. Pandola [b], H. Richter [d], A. A. Shikhin [a], W. A. Teasdale [c], E. P. Veretenkin [a], V. M. Vermul [a], J. F. Wilkerson [f] V. E. Yants [a], and G. T. Zatsepin [a]

[a]*Institute for Nuclear Research of the Russian Academy of Sciences, Moscow 117312, Russia*
[b]*INFN, Laboratori Nazionali del Gran Sasso (LNGS), S.S.17/bis Km 18+910, I-67010 L'Aquila, Italy*
[c]*Los Alamos National Laboratory, Los Alamos, NM 87545 USA*
[d]*Max-Planck-Institut für Kernphysik (MPIK), Postfach 103980, D-69029 Heidelberg, Germany*
[e]*Dipartimento di Fisica, Università di Milano 'La Bicocca' e INFN, Sezione di Milano, Via Emanueli, I-20126 Milano, Italy*
[f]*Department of Physics, University of Washington, Seattle, WA 98195 USA*

**Abstract**

We describe a cooperative measurement of the capture rate of solar neutrinos by the reaction $^{71}$Ga($\nu_e$, e$^-$)$^{71}$Ge. Extractions were made from a portion of the gallium target in the Russian-American Gallium Experiment SAGE and the extraction samples were transported to the Gran Sasso laboratory for synthesis and counting at the Gallium Neutrino Observatory GNO. Six extractions of this type were made and the resultant solar neutrino capture rate was $64^{+24}_{-22}$ SNU, which agrees well with the overall result of the gallium experiments. The major purpose of this experiment was to make it possible for SAGE to continue their regular schedule of monthly solar neutrino extractions without interruption while a separate experiment was underway to measure the response of $^{71}$Ga to neutrinos from an $^{37}$Ar source. As side benefits, this experiment proved the feasibility of long-distance sample transport in ultralow background radiochemical experiments and familiarized each group with the methods and techniques of the other.

## 1. Introduction

Two experiments, SAGE [1,2], at the Baksan Neutrino Observatory (BNO) in southern Russia, and GALLEX/GNO [3,4], at the Laboratori Nazionali del Gran Sasso (LNGS) in central Italy, have measured the capture rate of solar neutrinos on $^{71}$Ga during most of the last fifteen years. The GNO experiment ended extractions in April 2003, whereas the SAGE experiment is still continuing solar neutrino measurements.

From April through September 2004, SAGE conducted a new experiment to measure the response of $^{71}$Ga to an intense $^{37}$Ar neutrino source [5]. Since the samples from this experiment completely filled the SAGE counting system, it was not possible for SAGE to also count solar neu-trino samples at this time. This was the case even though the $^{37}$Ar experiment used only 26 tonnes of Ga and left 22 tonnes of Ga idle.

* Corresponding author.
  *E-mail address:* bclevela@surf.sno.laurentian.ca
  (B. T. Cleveland).

The GNO experiment at this time was, however, just finishing measurement of solar neutrino samples, and was able to make their counting system available to measure solar neutrino extractions from the SAGE gallium. Thus SAGE was able to continue their schedule of regular monthly solar neutrino extractions during the $^{37}$Ar experiment by bringing the final extracted solution to the LNGS for synthesis and counting by GNO. In this article we describe this collaborative experiment and give the results. To avoid any possible misconception, we wish to emphasize to the reader that since we separately used the customary methods and techniques of each of our experiments, extraction by SAGE and synthesis and counting by GNO, the measurements reported here do not constitute a check on the systematic effects of either experiment.

## 2. Experimental details

We give only a very brief description of experimental operations here. The reader is referred to [1,2] and [3,4], and the references therein, for full details on extraction and counting, respectively.

*Extraction.* Six extractions from the available Ga were made, with the first and last from the full amount of Ga, approximately 48 tonnes, and the other four extractions from 22 tonnes. The dates and other data for each extraction are given in Table 1. The efficiency of the extraction operations at Baksan was determined by adding to the Ga metal a known amount of Ge enriched in $^{76}$Ge to 94.6% as a Ga:Ge alloy and measuring the amount of Ge in the final extraction solution, whose volume was 100 ml. This measurement required 1 ml of solution and the concentration of Ge was determined by atomic-absorption spectroscopy. Problems were experienced in the secondary concentration steps for the first and last extractions and thus their efficiency was lower than the others. Inert natural Ge was added to the extraction solution to bring its total Ge content to ~10 $\mu$mol, the standard for GNO extractions, and the solution was then transported [1] to the LNGS as rapidly as possible.

*Synthesis and counter filling.* When the sample arrived at the LNGS, the extraction solution was put into a reaction flask, NaBH$_4$ was added to synthesize the counting gas GeH$_4$, the GeH$_4$ was separated from other gases by chromatography, and Xe was added to make the standard counting mixture of 30% GeH$_4$ and 70% Xe at ~750 mm Hg. The proportional counter was then filled, the counter was calibrated, and measurement began. All these steps (and the extraction procedure) were done rapidly so that the total time from the end of solar neutrino exposure to the start of counting, including transport from Baksan to Gran Sasso, averaged 61.8 hours for the six extractions. The efficiency of synthesis was (96–99)% for all extractions except one (S005).

*Counting.* The extraction samples were measured in GNO ultralow background counters which had previously been used to measure 11 of the last 13 GNO solar neutrino extractions. The counting system recorded the event energy and the pulse waveform for 400 ns after onset with a time resolution of 0.2 ns/point. Two fast transient digitizers were used on each counting channel, one for the $L$ peak and the other for the $K$ peak. The energy response of each counter was calibrated with an x-ray source at the start of counting and then approximately each month until counting ended. The information on the counters used and the times of acquisition is given in Table 2.

*Time cuts.* Two time cuts were made to the data to reduce the influence of $^{222}$Rn. These cuts are necessary as some of the $\beta$ particles or recoil nuclei in the $^{222}$Rn decay chain can produce pulses that are indistinguishable from true $^{71}$Ge events.

Fortunately, the decays at the beginning of the chain, $^{222}$Rn to $^{218}$Po and $^{218}$Po to $^{214}$Pb, give $\alpha$ particles whose pulses usually saturate the counter's energy scale, indicating that a $^{222}$Rn decay may have occurred. Because the chain usually takes about an hour to complete, the first time cut thus deleted all counting time from 15 min before to

---

[1] Hand carried via airports at Mineralnye Vody (Russia), Moscow, and Rome.

Table 1
Extraction schedule and related parameters. Two names are given for each extraction, the first was used at Baksan and the second at LNGS. The times of exposure are given in days of year 2004. The overall extraction efficiency is in the column headed "into GeH$_4$." The final extraction was made from all available Ga which consisted of three batches with different exposure times. The separate exposures of the various Ga batches were taken into account for this extraction by adjusting the effective total mass assuming constant $^{71}$Ge production rate.

| Extraction name | Extraction date (2004) | Exposure time | | Mass Ga (tonnes) | $^{76}$Ge carrier mass ($\mu$mol) | Added nat. Ge std. sol. ($\mu$mol) | Extraction efficiency | |
|---|---|---|---|---|---|---|---|---|
| | | Begin | End | | | | into extract | into GeH$_4$ |
| 7R1044 = S001 | 22 Apr | 90.16 | 113.76 | 48.287 | 3.50 | 7.95 | 0.65 | 0.63 |
| 3R1054 = S002 | 16 May | 114.18 | 137.58 | 22.028 | 3.62 | 7.95 | 0.92 | 0.88 |
| 3R1064 = S003 | 24 Jun | 137.89 | 176.39 | 22.001 | 3.74 | 7.68 | 0.92 | 0.89 |
| 3R1074 = S004 | 25 Jul | 182.46 | 207.36 | 21.953 | 3.50 | 7.68 | 0.98 | 0.96 |
| 3R1084 = S005 | 23 Aug | 207.53 | 236.83 | 21.929 | 3.55 | 7.68 | 1.01 | 0.91 |
| 7R1094 = S006 | 25 Sep | 236.95 | 269.85 | 42.420 | 8.07 | 3.29 | 0.76 | 0.75 |

Table 2
Counting parameters. $\Delta$ is the exponentially weighted live time starting from the time of the end of extraction. The live time and $\Delta$ include all time cuts.

| Extraction name | Counter name | Counting efficiency after energy and pulse shape cuts | | Day counting began in 2004 | Live time of counting (days) | $\Delta$ |
|---|---|---|---|---|---|---|
| | | $L$ peak | $K$ peak | | | |
| S001 | SC136 | 0.329 | 0.381 | 116.71 | 222.9 | 0.811 |
| S002 | SI106 | 0.309 | 0.314 | 139.90 | 200.7 | 0.853 |
| S003 | SI108 | 0.290 | 0.316 | 178.66 | 151.7 | 0.733 |
| S004 | FC093 | 0.294 | 0.341 | 209.72 | 163.3 | 0.842 |
| S005 | FC174 | 0.297 | 0.330 | 239.54 | 130.5 | 0.744 |
| S006 | FC126 | 0.303 | 0.321 | 272.70 | 101.2 | 0.794 |

180 min after every event with saturated energy. This cut removed most of these false $^{71}$Ge events.

As a further defense against $^{222}$Rn, the counting system contained a digitizer with 800-$\mu$s time span which was able to recognize the delayed coincidence between the $\beta$ decay of $^{214}$Bi, the daughter of $^{214}$Pb, and the subsequent $\alpha$ decay of $^{214}$Po, whose half-life is 164 $\mu$s. If such an event was detected, called a BiPo event, a second time cut was made which deleted all counting time from 180 min prior to the time of the BiPo event up to the time of the event. Further, since the BiPo signified the completion of the $^{222}$Rn decay chain at 22-y $^{210}$Pb, if a time cut of the first type was in progress when the BiPo event was detected, the first cut was ended at the time of the BiPo.

Although $^{222}$Rn mainly enters the counter when it is filled and thus has the greatest influence at the start of counting, some $^{222}$Rn may enter the counting gas at any time from the decay of $^{226}$Ra in the counter materials, and thus these two time cuts were applied for the entire data acquisition period.

*Event selection.* Candidate $^{71}$Ge events were selected by making cuts on both energy and pulse shape. The energy calibration was used to predict the peak positions and widths of the $^{71}$Ge $L$ and $K$ peaks and candidate events were required to have an energy within ±1 full width at half maximum of the predicted peak position.

The cut on pulse shape was made by fitting the pulses with a semi-empirical function and then feeding the fit parameters into a neural net [6]. The net was trained on reference pulses acquired with $^{71}$Ge-filled counters and on background-type pulses produced by $\gamma$ rays from a $^{137}$Cs source. This cut should reject 70% of background events in the $L$- and $K$-peak regions.

The counting efficiency after making these energy and pulse-shape cuts is given in Table 2. The absolute efficiency of all counters has been directly measured with $^{69}$Ge.

*Time analysis.* These selection criteria yielded a set of events from each extraction in both the $L$- and $K$-peak regions which were candidate $^{71}$Ge de-cays. The final step of the experimental procedures was to estimate the solar neutrino production rate. This was done by assuming the set of candidate $^{71}$Ge events was produced by the superposition of events from $^{71}$Ge, which decay with an exponential distribution with an 11.43-d half-life, plus background events whose distribution was constant in time. A maximum-likelihood method [7] was used to decompose the sequence of event times into these two components and thus to determine the most likely number of $^{71}$Ge decays in the data set. Combining this datum with the counting efficiency and the counting times yields the best estimate for the $^{71}$Ge production rate.

For each extraction this procedure was done separately for the events in the $L$ peak and in the $K$ peak and the results are given for each extraction in Table 3. The overall likelihood function for each extraction was defined as the product of the separate likelihood functions for the $L$- and $K$-peak regions, where the background rates in each peak were independent. The best fit production rate was found by maximizing this function, allowing the independent background rates in the $L$ and $K$ peaks to be free variables. The combined $L + K$-peak results are given for each extraction in Table 4. The overall likelihood function for the entire experiment and the overall best fit production rate were defined and calculated in an analogous manner. The uncertainties were calculated as described in [1].

Also given in these tables are the values of a parameter that tests the goodness of fit of the sequence of event times to the assumed model for the time distribution. The probabilities for the combinations are close to 50%, indicating the model fits the data quite well.

Time analysis programs were independently developed by SAGE and by GALLEX/GNO. These two programs gave the same results within round-off error, except for those single extractions for which the best-fit number of $^{71}$Ge atoms was close to zero. In this case different results were obtained because the SAGE analysis restricts the background and production rates to positive values, whereas the GALLEX/GNO analysis permits these quantities to become negative. The values given in the tables are when positivity is required.

If negative background and production rates are allowed, there is no unique method of analysis, and the results depend on the particular conditions imposed on the likelihood function as one extends to negative values outside the physical region. The choice made by GALLEX/GNO is to require that this function be positive at the times of occurrence of all observed events. For the combination of $L$ and $K$ peaks given in Table 4, the only runs for which different results are obtained are S006, for which the result is $-18^{+37}_{-28}$ SNU, and S004, for which the negative uncertainty extends to −60 SNU. The overall combined result is the same with both methods of analysis.

## 3. Results

Before giving the final results, we must correct for two additional effects:

- the Earth-Sun distance variation. Since the BNO-LNGS measurements were made when the Sun was near aphelion, this correction increased the capture rate by (1–3)%.
- the effect of the $^{37}$Ar source, which was present during all extractions except the first. The chemical reactors that contained the Ga used in BNO-LNGS were located (4–12) m distant from the source, but the effect of the source was still not negligible. Based on the source strength, the cross section for capture of $^{37}$Ar neutrinos by $^{71}$Ga [10], and accurate knowledge of the reactor shape, location, and Ga content, the number of $^{71}$Ge atoms expected to be produced by the source in each reactor was calculated and converted into an equivalent constant solar neutrino rate [11]. This effective rate is given in Table 4 and has been subtracted from the total production rate for each extraction.

The solar neutrino capture rates in Tables 3 and 4 have been corrected for these two effects. The overall result of the BNO-LNGS experiment is the $L+K$ combination in Table 4: $64^{+24}_{-21}$ SNU [2], where the error is only statistical.

---

[2] 1 SNU = 1 neutrino capture/s in a target that contains $10^{36}$ atoms of the neutrino absorbing isotope.

Table 3
Results of separate analysis of $L$-peak and $K$-peak events for each extraction. The SNU rates are corrected for the Earth-Sun distance variation and the fixed background from the $^{37}$Ar source. Errors are statistical with 68% confidence. The parameter $Nw^2$ measures the goodness of fit of the sequence of event times to the model assumed in analysis [8,9]. The probability was inferred from $Nw^2$ by simulation and has an uncertainty of $\sim\pm 2\%$.

| Extraction name | Number of candidate events | | Number fit to $^{71}$Ge | | Solar neutrino capture rate (SNU) | | $Nw^2$ | | Probability (percent) | |
|---|---|---|---|---|---|---|---|---|---|---|
| | L | K | L | K | L | K | L | K | L | K |
| S001 | 12 | 7 | 4.3 | 0.6 | $143^{+97}_{-72}$ | $16^{+46}_{-16}$ | 0.085 | 0.028 | 33 | 97 |
| S002 | 3 | 6 | 1.5 | 2.2 | $61^{+96}_{-59}$ | $98^{+113}_{-78}$ | 0.064 | 0.048 | 63 | 72 |
| S003 | 6 | 4 | 4.1 | 0.2 | $218^{+153}_{-114}$ | $0^{+53}_{-0}$ | 0.069 | 0.076 | 54 | 61 |
| S004 | 9 | 5 | 1.2 | 0.8 | $58^{+123}_{-58}$ | $28^{+103}_{-28}$ | 0.044 | 0.065 | 80 | 60 |
| S005 | 8 | 4 | 0.1 | 4.0 | $0^{+76}_{-0}$ | $205^{+121}_{-87}$ | 0.034 | 0.080 | 93 | 68 |
| S006 | 8 | 6 | 0.2 | 0.0 | $5^{+59}_{-5}$ | $0^{+29}_{-0}$ | 0.112 | 0.053 | 37 | 75 |
| Combined | 46 | 32 | 11.6 | 8.0 | $81^{+37}_{-32}$ | $49^{+32}_{-27}$ | 0.056 | 0.046 | 62 | 70 |

Table 4
Results of combined analysis of $L$- and $K$-peak events for each extraction. See caption for Table 3 for further explanation.

| Extraction name | Number of candidate events | Number fit to $^{71}$Ge | Fixed rate by $^{37}$Ar source (SNU) | Solar neutrino capture rate (SNU) | $Nw^2$ | Probability (percent) |
|---|---|---|---|---|---|---|
| S001 | 19 | 4.3 | 0.0 | $67^{+48}_{-37}$ | 0.076 | 43 |
| S002 | 9 | 3.6 | 18.1 | $79^{+70}_{-52}$ | 0.076 | 45 |
| S003 | 10 | 4.1 | 11.4 | $99^{+84}_{-64}$ | 0.091 | 34 |
| S004 | 14 | 2.0 | 5.8 | $43^{+75}_{-43}$ | 0.051 | 74 |
| S005 | 12 | 5.2 | 3.3 | $139^{+78}_{-59}$ | 0.048 | 79 |
| S006 | 14 | 0.1 | 1.4 | $0^{+25}_{-0}$ | 0.102 | 46 |
| Combined | 78 | 19.6 | | $64^{+24}_{-21}$ | 0.045 | 77 |

The total number of events that time analysis assigned to $^{71}$Ge in these six extractions was 19.6. When the number of events detected is quite low, as was the case in this experiment, the time cuts to reduce the influence of $^{222}$Rn are quite essential. If these cuts are not applied, it is found that the total number of candidate events in the $L+K$ combination increases from 78 to 91. Moreover, since most $^{222}$Rn is expected to enter the counter when it is filled and its half-life is 3.8-d, all of these additional 13 events occurred in the first 15 d of counting so that the number of events assigned to $^{71}$Ge becomes 36.2, and the apparent capture rate nearly doubles to $123^{+27}_{-25}$ SNU. In a separate experiment, the GNO group has measured the efficiency of these time cuts to reject false $^{71}$Ge events to be 100% with an uncertainty of 4.4%. Since the result with the cuts is 59 ± 35 SNU less than without, we set -2.6 SNU as the systematic uncertainty associated with $^{222}$Rn.

This value and the other components of the systematic uncertainty are listed in Table 5. The items for extraction, fast neutrons underground, $^{232}$Th, $^{226}$Ra, and cosmic-ray muons are the standard values for SAGE, as described in [1]. The items for synthesis, event selection, and internal $^{69}$Ge are the standard values for GNO, as given in [4,6]. The uncertainty in the subtraction for the $^{37}$Ar source was dominated by the cross section uncertainty of $^{+7}_{-3}$%. It was determined by setting the cross section at these two extremes, re-evaluating the fixed rate from the $^{37}$Ar source, and re-calculating the solar neutrino rate.

Two systematic errors that are not present in normal solar neutrino runs, but were introduced by the sample transport, are due to cosmic-ray neutrons:
- production of $^{71}$Ge on the Ge isotopes in the carrier with mass greater than 71 by spallation induced by high-energy neutrons. We can estimate the magnitude of this effect by convoluting the

Table 5
Summary of the contributions to the systematic uncertainty in the measured neutrino capture rate. The SNU values for extraction and synthesis efficiency are based on a rate of 64 SNU. The total is taken to be the quadratic sum of the individual contributions. For comparison, the statistical uncertainty in the result of the BNO-LNGS experiment is $^{+24}_{-21}$ SNU.

| Origin of uncertainty | Uncertainty in % | Uncertainty in SNU |
|---|---|---|
| Chemical extraction efficiency | | |
|     Mass of gallium | ±0.3 | ±0.2 |
|     Mass of added Ge carrier | ±2.1 | ±1.3 |
|     Carrier carryover | ±0.8 | ±0.5 |
|     Chemical extraction subtotal | ±2.3 | ±1.4 |
| Synthesis efficiency | ±2.0 | ±1.3 |
| Counting efficiency | | |
|     Event selection in energy | | ±1.3 |
|     Event selection by neural net | | ±0.9 |
|     Counting efficiency subtotal | | ±1.6 |
| Background events that mimic $^{71}$Ge | | |
|     Internal $^{222}$Rn | | <−2.6 |
|     Internal $^{69}$Ge | | <−1.0 |
|     Background events subtotal | | <−2.8 |
| Nonsolar neutrino production of $^{71}$Ge | | |
|     $^{37}$Ar source | | +0.18, −0.42 |
|     Fast neutrons underground | | <−0.02 |
|     Fast neutrons during transit | | <−0.19 |
|     Thermal neutrons during transit | | <−0.21 |
|     $^{232}$Th | | <−0.04 |
|     $^{226}$Ra | | <−0.7 |
|     Cosmic-ray muons | | <−0.7 |
|     Nonsolar subtotal | | +0.18, <−1.1 |
| Total systematic uncertainty | | +2.5, −3.9 |
cross sections for the various reactions with the measured neutron flux. There do not appear to be any measurements or calculations of the spallation cross sections on the Ge isotopes to $^{71}$Ge, but we can approximate them by the analogous calculations for the reactions that yield $^{68}$Ge, which have been calculated for $\beta\beta$ experiments with Ge [12,13]. Integrating from threshold up to 1000 MeV, and using the high-energy neutron spectrum from the review of Ziegler [14] and the number of Ge isotopes inferred from Table 1, gives a production rate of $\sim 10^{-4}$ $^{71}$Ge produced/(d run) at sea level. The Ge sample spent $\sim$2 d close to sea level, and about 0.2 d during the airplane flights from Mineralnye Vody to Rome at $\sim$10 km altitude, where the neutron flux is a factor of $\sim$200 times higher [14]. Adding these components, and using the reduced natural Ge concentration in run S006 (see Table 1), gives a total production of $\sim$0.021 atoms of $^{71}$Ge in the sum of all six runs. Considering the large uncertainties in the various factors that enter the production rate, the error in this estimate is approximately a factor of five.

- production of $^{71}$Ge by the capture of thermal neutrons in the (n, $\gamma$) reaction on $^{70}$Ge. The capture cross section for this reaction, including the contribution to the 198-keV level which decays to the ground state, is 3.4 barns. Using the number of $^{70}$Ge atoms in each run, the thermal neutron flux at sea level of $\sim$4 neutrons/(cm$^2$ hr) [15], the fact that the thermal neutron flux at high altitude is $\sim$100 times that at sea level, and assuming the same exposure times as for the high-energy neutron spallation considered above, we estimate a total of 0.038 atoms of $^{71}$Ge produced by thermal neutrons in the six BNO-LNGS runs. The error of this calculation is about a factor of three, mainly due to the uncertain neutron flux which depends on the water content of the air.

For comparison, assuming the solar neutrino capture rate is the result of the BNO-LNGS experiment, 64 SNU, then the number of $^{71}$Ge atoms produced in 22 tonnes of Ga and present at the end of a 30-d exposure is 5.8. For the six BNO-LNGS runs this leads to an expected total of 35 atoms of $^{71}$Ge present at the time of extraction. Upper limits for these systematic uncertainties are thus 0.19 SNU for high-energy neutrons and 0.21 SNU for thermal neutrons.

Another item to note is that the rate of background events in the BNO-LNGS extractions was 0.06 events/d in the sum of the $L$ and $K$ peaks, which is identical to the rate for extractions from the GNO tank of GaCl$_3$:HCl solution. The rate of saturated events was also about the same in BNO-LNGS as in GNO.

## 4. Summary and conclusions

The six solar neutrino runs that were extracted at BNO and transported to LNGS for synthesis and counting yield a combined best-fit result of

$64^{+24}_{-21}$ (stat) $^{+2.5}_{-3.9}$ (syst) SNU. Combining the uncertainties in quadrature gives $64^{+24}_{-22}$ SNU. This result is in good agreement with the combined result from 244 solar neutrino extractions in the SAGE and GALLEX/GNO experiments, which is 68.1 ± 3.75 SNU [16].

The results of these six measurements in the BNO-LNGS experiment will be combined with the previous (and ongoing) measurements by SAGE so as to improve our overall knowledge of the solar neutrino capture rate in $^{71}$Ga.

The transport of our radiochemical samples from Baksan to Gran Sasso proved to be entirely practical. Furthermore, no effects associated with the sample transport were identified that appreciably degrade the systematic uncertainty. We are thus confident that if this experiment were continued for many more runs, or extended to larger masses of gallium, the solar neutrino capture rate would be measured with high precision.

Acknowledgements

We thank the members of GNO who did not directly participate in this experiment for their cooperation. The SAGE group wishes to thank V. A. Matveev for his constant fruitful attention to the experiment and acknowledges award no. 02-02-16776 from the Russian Foundation for Basic Research and grant no. NS-1782.2003.2 from the President of the Russian Federation for the Leading Scientific Schools of Russia. This work was also supported by the European Community through a grant from the Transnational Access to Research Infrastructure (TARI) program (project no. P02/04 at LNGS).

References

[1] J. N. Abdurashitov, V. N. Gavrin, S. V. Girin, V. V. Gorbachev, T. V. Ibragimova, A. V. Kalikhov, N. G. Khairnasov, T. V. Knodel, I. N. Mirmov, A. A. Shikhin, E. P. Veretenkin, V. M. Vermul, V. E. Yants, G. T. Zatsepin, T. J. Bowles, W. A. Teasdale, D. L. Wark, M. L. Cherry, J. S. Nico, B. T. Cleveland, R. Davis, Jr., K. Lande, P. S. Wildenhain, S. R. Elliott, and J. F. Wilkerson, Phys. Rev. C 60 (1999) 055801 [arXiv: astro-ph/9907113].

[2] J. N. Abdurashitov, E. P. Veretenkin, V. M. Vermul, V. N. Gavrin, S. V. Girin, V. V. Gorbachev, P. P. Gurkina, G. T. Zatsepin, T. V. Ibragimova, A. V. Kalikhov, T. V. Knodel, I. N. Mirmov, N. G. Khairnasov, A. A. Shikhin, V. E. Yants, T. J. Bowles, W. A. Teasdale, J. S. Nico, J. F. Wilkerson, B. T. Cleveland, and S. R. Elliott, J. Exp. Theor. Phys. 95 (2002) 181, Zh. Eksp. Teor. Fiz. 122 (2002) 211 [arXiv:astro-ph/0204245].

[3] W. Hampel, J. Handt, G. Heusser, J. Kiko, T. Kirsten, M. Laubenstein, E. Pernicka, W. Rau, M. Wojcik, Y. Zakharov, R. v. Ammon, K. H. Ebert, T. Fritsch, E. Henrich, L. Stielglitz, F. Weirich, M. Balata, M. Sann, F. X. Hartmann, E. Bellotti, C. Cattadori, O. Cremonesi, N. Ferrari, E. Fiorini, L. Zanotti, M. Altmann, F. v. Feilitzsch, R. Mößbauer, S. Wänninger, G. Berthomieu, E. Schatzmann, I. Carmi, I. Dostrovsky, C. Bacci, P. Belli, R. Bernabei, S. d'Angelo, L. Paoluzi, M. Cribier, J. Rich, M. Spiro, C. Tao, D. Vignaud, J. Boger, R. L. Hahn, J. K. Rowley, R. W. Stoenner, and J. Weneser, Phys. Lett. B 447 (1999) 127.

[4] M. Altmann, M. Balata, P. Belli, E. Bellotti, R. Bernabei, E. Burkert, C. Cattadori, R. Cerulli, M. Chiarini, M. Cribier, S. d'Angelo, G. Del Re, K. H. Ebert, F. v. Feilitzsch, N. Ferrari, W, Hampel, F. X. Hartmann, E. Henrich, G. Heusser, F. Kaether, J. Kiko, T. Kirsten, T. Lachenmaier, J. Lanfranchi, M. Laubenstein, K. Lützenkirchen, K. Mayer, P. Moegel, D. Motta, S. Nisi, J. Oehm, L. Pandola, F. Petricca, W. Potzel, H. Richter, S. Schoenert, M. Wallenius, M. Wojcik, and L. Zanotti, Phys. Lett. B 616 (2005) 174 [arXiv: hep-ex/0504037].

[5] J. N. Abdurashitov, V. N. Gavrin, S. V. Girin, V. V. Gorbachev, P. P. Gurkina, T. V. Ibragimova, A. V. Kalikhov, N. G. Khairnasov, T. V. Knodel, V. A. Matveev, I. N. Mirmov, A. A. Shikhin, E. P. Veretenkin, V. M. Vermul, V. E. Yants, G. T. Zatsepin, T. J. Bowles, S. R. Elliott, W. A. Teasdale, J. S. Nico, B. T. Cleveland, W. C. Haxton, J. F. Wilkerson, A. Suzuki, K. Lande, Yu. S. Khomyakov, V. M. Poplavsky, V. V. Popov, O. V. Mishin, A. N. Petrov, B. A. Vasiliev, S. A. Voronov, A. I. Karpenko, V. V. Maltsev, N. N. Oshkanov, A. M. Tuchkov, V. I. Barsanov, A. A. Janelidze, A. V. Korenkova, N. A. Kotelnikov, S. Yu. Markov, V. V. Selin, Z. N. Shakirov, A. A. Zamyatina, and S. B. Zlokazov, Proc. XI International Wokshop on "Neutrino Telescopes", Venice, Italy, 22–25 February 2005.

[6] L. Pandola, C. Cattadori, and N. Ferrari, Nucl. Instrum. Methods Phys. Res. A 522 (2004) 521.

[7] B. T. Cleveland, Nucl. Instrum. Methods Phys. Res. 214 (1983) 451.

[8] A. W. Marshall, Ann. Math. Stat. 29 (1958) 307.

[9] B. T. Cleveland, Nucl. Instrum. Methods Phys. Res. A 416 (1998) 405.

[10] J. N. Bahcall, Phys. Rev. C 56 (1997) 3391 [arXiv: hep-ph/9710491].

[11] F. Kaether, *BNO-LNGS source background*, MPI internal note, 15 February 2005.

[12] S. Cebrián, Rev. Real Academia de Ciencias. Zaragoza 59 (2004) 7.

[13] Majorana collaboration, *White paper on the Majorana zero-neutrino double-beta decay experiment*, arXiv:nucl-ex/0311013 (2003).

8[14] J. F. Ziegler, IBM J. of Res. and Develop. 42 (1998) 117.

[15] J. D. Dirk, M. E. Nelson, J. F. Ziegler, A. Thompson, and T. H. Zabel, IEEE Trans. on Nucl. Sci. 50 (2003) 2060.

[16] C. Cattadori, N. Ferrari, and L. Pandola, Nucl. Phys. B (Proc. Suppl.) 143 (2005) 3.